\documentclass[acmlarge]{acmart}
\makeatletter
\newcommand{\myconfshort}{\acmConference@shortname}
\newcommand{\myconffull}{\acmConference@name}
\newcommand{\myconfdate}{\acmConference@date}
\newcommand{\myconfloc}{\acmConference@venue}
\AtBeginDocument{
  \fancypagestyle{firstpagestyle}{
    \fancyhead{}%
    \fancyfoot[C]{}%
  }
  \fancyhf{}
  \fancyhead[LO]{\@headfootfont\shorttitle}%
  \fancyhead[RE]{\@headfootfont\@shortauthors}%
  \fancyhead[LE]{\@headfootfont\footnotesize \myconfshort, \myconfdate, \myconfloc}%
  \fancyhead[RO]{\@headfootfont\footnotesize \myconfshort, \myconfdate, \myconfloc}%
  \fancyfoot[C]{}%
}
\makeatother
\acmBooktitle{\conffull\@ (\confshort), \confdate, \confloc}

\copyrightyear{2026}
\acmYear{2026}
\setcopyright{cc}
\setcctype{by}
\acmConference[FAccT '26]{The 2026 ACM Conference on Fairness, Accountability, and Transparency}{June 25--28, 2026}{Montreal, QC, Canada}
\acmBooktitle{The 2026 ACM Conference on Fairness, Accountability, and Transparency (FAccT '26), June 25--28, 2026, Montreal, QC, Canada}
\acmDOI{10.1145/3805689.3812375}
\acmISBN{979-8-4007-2596-8/2026/06}

\usepackage{subcaption}
\usepackage{amsfonts}
\usepackage{tabularx}
\usepackage{url}
\usepackage{xcolor}
\usepackage{ulem}
\AtBeginDocument{%
  }

\begin{document}

\title[Accounting for Missing Demographic Data when Mitigating Ad Delivery Skew]{Into the Unknown: Accounting for Missing Demographic Data when Mitigating Ad Delivery Skew}

\author{Isabel Corpus}
\email{isc36@cornell.edu}
\orcid{}
\affiliation{%
  \institution{Cornell University}
  \city{New York}
  \state{New York}
  \country{USA}
}

\author{Allison Koenecke}
\email{koenecke@cornell.edu}
\affiliation{%
  \institution{Cornell University}
  \city{New York}
  \state{New York}
  \country{USA}}

\begin{abstract}
Online advertising platforms use algorithmic systems to power the process of matching ads to users, termed \textit{ad delivery}.
Prior audits have demonstrated that ad delivery can be \textit{skewed} by demographic attributes, such that ads are systematically under-delivered to certain groups despite advertiser intent to reach groups proportionally.
This under-delivery raises a serious concern in the context of ads promoting public services, which might prevent certain groups of individuals from accessing information about resources on the basis of their demographic identity. 
In the absence of platform-provided solutions to skewed ad delivery, advertisers can counteract skew by targeting demographic groups directly. 
However, direct targeting excludes users whose demographics the platform cannot infer (``unknown users'') if advertising platforms do not provide a way to target unknown users directly, as is the case on Google Ads.
We collaborate with a state-level government agency to reduce gender-based skew in ad delivery with an intervention that accounts for unknown users while incorporating gender-based targeting.
In particular, we design a budget split intervention that directly incorporates unknown users and targets users with Google-inferred gender labels (i.e., male, female).
We find that this intervention is a valuable approach to addressing ad delivery skew without excluding unknown users, and serves as a middle ground in the trade-off between higher costs (from more granular demographic targeting) and skew (from ignoring demographics entirely).
This approach is responsive to the needs of real-world, resource-constrained advertisers who are committed to the equitable distribution of public service outreach via online advertising.
We conclude with recommendations for government advertisers, online advertising platforms, and researchers. 
\end{abstract}



\maketitle

\section{Introduction}

The success of online advertising platforms is attributable, in part, to the algorithms that platforms use to maximize advertiser and consumer utility. 
However, these systems can introduce bias in the \textit{ad delivery} stage of the advertising optimization process, when ads are programmatically allocated to users~\cite{shuai2013rtb}. 
Researchers have identified \textit{skew} in ad delivery, whereby ads are disproportionately delivered to particular demographic groups, despite advertiser intent to reach demographic groups proportionally~\cite{ali2019discrimination}.
Skewed ad delivery is an expected feature of online advertising platforms, which use predictions of ad relevance and quality to route ads toward the users who are most likely to engage with the content. 
These predictions are expected to increase consumer utility by showing users relevant content and increase advertiser utility by driving up ad engagement~\cite{ayman2012, yan2009}. 
However, predictions of relevance may correlate with sensitive attributes, skewing delivery toward particular groups~\cite{ali2019discrimination}.

In domains such as housing, employment, and credit, demographic skew in ad delivery can amount to discrimination~\cite{ali2019discrimination}.
While not covered by anti-discrimination law, skewed ad delivery for public goods raises similar concerns: it systematically undermines initiatives intended to reduce social disparities and contributes to allocative harms by inequitably distributing information~\cite{suresh2021framework}. 
For example, researchers found that a California ad campaign for a nutritional public benefit program consistently under-delivered ads to Spanish-speakers, which could exacerbate gaps in resource attainment~\cite{koenecke2023popular}.
Despite the consequences, public sector advertising has been understudied relative to contexts that are covered by American anti-discrimination law~\cite{imana2025externalevaluation, imana2021jobads} or that explicitly encode a stereotype threat, like STEM education~\cite{lambrecht2019algorithmic, ali2019discrimination}. 
In this study, we focus on the relatively understudied domain of public sector advertising in collaboration with a state-level government agency.

To isolate the causes of skewed ad delivery, researchers have developed rigorous auditing methods~\cite{imana2025externalevaluation, lambrecht2019algorithmic, ali2019discrimination}. 
However, the careful methodological construction of these audits presents substantial financial and operational costs, thus hindering their adoption for the evaluation of real-world advertising campaigns.
One key gap is in sourcing demographic data: prior external audits have relied on voter records~\cite{imana2025externalevaluation}, whereas advertisers often use platform-inferred demographics.
Although voter records provide a reliable ground truth, they are not broadly available. Alternative data collection processes may be unrealistic for resource-constrained government agencies that outsource advertising operations to private vendors.
We present an evaluation and intervention for skewed ad delivery that uses platform-inferred demographic labels, and is thus responsive to the needs of government advertisers in practice.

We partnered with a state-level government agency to evaluate and address skewed ad delivery in the context of a real-world Google Ads advertising initiative.
The initiative advertised a government website with information about taxes, supportive services, and grants for entrepreneurs or people planning to start businesses. 
Historically, women and racial minorities were underrepresented among business owners in the state. 
The government agency sought to prevent worsening disparities in business ownership by ensuring the ad initiative reached Google users at a rate proportional to state population demographics, with a particular focus on gender.

Our evaluation revealed consistent skew in ad delivery toward users labeled male by Google.
To address this skew, we designed and implemented a budget split intervention.
Budget split interventions split a single ad campaign into multiple new campaigns that each apply group-specific targeting, thereby increasing advertiser control over the distribution of ad delivery (see details in Section~\ref{sec:budget split definition}).
Unlike prior work, our budget split intervention considered ``unknown users,'' for whom Google fails to assign demographic labels.
By including unknown users in our targeting, we avoid excluding them from potentially receiving public resource outreach.
This is important for public interest advertising, as unknown users may be more likely to have marginalized identities (see Section~\ref{sec:advertising tradeoffs}).
We found that unknown users reflect a necessity of advertising on Google Search and present an opportunity for mitigating ad delivery skew: targeting unknown users can reduce skew in ad delivery while ameliorating the cost trade-off of budget split interventions.

\textbf{We asked: (1) Has ad delivery in this government advertising campaign historically shown gender-based skew? (2) If so, how can we design a budget split intervention that enforces parity in ad delivery, minimizes utility trade-offs, and includes unknown users? (3) How effective and cost-efficient is our intervention?
}

Our work builds on and contributes to the FAccT literature on fairness in advertising (e.g.,~\cite{imanaAuditingEducationAds2024, andric2023reconcilinggovernmental, imana2025externalevaluation, imana2025inferreddemographics, baumann2024FairnessInOnlineAdDelivery, sampson2023queerpeoples}) by raising awareness of the importance of unknown users, demonstrating the fairness-privacy and fairness-utility trade-offs facing public interest advertisers, and showing evidence that an unknown-aware budget split intervention can ameliorate these trade-offs.
Concretely, we make three contributions to the fairness in ad delivery literature: (1) we identify persistent ad delivery skew in a case study of a real-world government advertising campaign, (2) to address this skew, we design and implement a budget split intervention that includes users with unknown demographic labels, and (3) we show this intervention reduces skew while mitigating cost trade-offs.
We use the findings from our case study to present recommendations for government advertisers, ad delivery platforms, and researchers.

\section{Background and Related Work}
\subsection{External Auditing of Skewed Ad Delivery}
\label{subsection: auditing skewed ad delivery}
Skewed ad delivery occurs when ads are shown disproportionately to particular user subgroups, despite advertiser intent to reach a demographically balanced audience~\cite{ali2019discrimination}. 
Prior auditing research has demonstrated that ad delivery algorithms are responsible for introducing skew in ad delivery in high-risk contexts like employment and education~\cite{ali2019discrimination,
imana2021jobads, imanaAuditingEducationAds2024}.
In 2013,~\citet{sweeneyDiscriminationOnlineAd2013} first demonstrated that Google AdSense was more likely to deliver ads for criminal arrest records for Black names than white names.
Since then, researchers have demonstrated persistent, systematic ad delivery skew against historically subordinated groups in high-impact domains like education, employment, and housing~\cite{speicherPotentialDiscriminationOnline2018, ali2019discrimination, sapiezynskiAlgorithmsThatDont2022, lambrechtApparentAlgorithmicDiscrimination2024, lambrecht2019algorithmic, imana2021jobads, imanaAuditingEducationAds2024, aliAdDeliveryAlgorithms2021}. 
Furthermore, the direction of skewed ad delivery exacerbates stereotypes and social biases. 
For example, ads about STEM and software engineer jobs were more likely to be delivered to male users than female users~\cite{lambrecht2019algorithmic}; ads for predatory colleges were more likely to be delivered to Black than white users~\cite{imanaAuditingEducationAds2024}. 

Three of the core factors that drive skewed ad delivery are (1) predicted relevance, (2) competition, and (3) baseline engagement rates~\cite{sapiezynskiAlgorithmsThatDont2022, baumann2024FairnessInOnlineAdDelivery,
ali2019discrimination}.
First, predicted relevance scores are intended to benefit both consumers, by showing them relevant content and filtering out click-spam ads, and advertisers, by increasing likelihood of ad engagement~\cite{meta_business_help_ad_delivery_2025, vacha2013spamads}.
However, predicted relevance scores might rely on stereotypes; as demonstrated by \citet{ali2019discrimination}, bodybuilding advertisements on Facebook were delivered to over 80\% male users. 
Second, competition increases when particular subgroups of users are more desired by advertisers or less available on the platform, which drives up the cost of reaching those users~\cite{dwork2018FairnessUnderComposition, liu2014OSNAdAuctions, lambrecht2019algorithmic,saeztrumper2014beyondcpmandcpc}. 
As advertising platforms seek to optimize for cost-effectiveness, advertisers with limited budgets may then fail to reach expensive users, causing skewed ad delivery toward the affordable user groups.
\citet{lambrecht2019algorithmic} demonstrate this for STEM job ads that were intended to be gender-neutral, but were over-delivered to men, because women were more competitive and thus expensive to reach.
Third, baseline engagement rates refer to the different distributions of engagement within user subgroups~\cite{baumann2024FairnessInOnlineAdDelivery}.
As noted by~\citet{baumann2024FairnessInOnlineAdDelivery}, differing baseline engagement implies that skewed ad delivery benefits advertising utility: enforcing statistical parity in ad delivery under unequal base engagement rates would degrade system performance, even under a perfectly accurate classifier. 
Although these factors contribute to utility for advertisers and users, they also systematically prevent users of certain demographic groups from accessing the information and opportunities in government outreach campaigns~\cite{imana2025externalevaluation, ali2019discrimination, lambrecht2019algorithmic}.

We extend prior audit literature by evaluating ad delivery skew in an active real-world advertising campaign promoting resources for small business owners.
We find that, in some contexts, ads had been over-delivered to users with the Google-inferred male label. 
In Section 3 we describe the method and results of our evaluation.
In Section 4 we discuss the intervention we design to reduce this skew in ad delivery.

\subsection{Budget Split Intervention to Address Skewed Ad Delivery}
\label{sec:budget split definition}
Advertising platforms do not permit U.S. advertisers to apply demographic targeting for sensitive topics (housing, credit, and employment). 
However, these controls do not counter the inherent risk of skewed ad delivery introduced through predicted ad relevance, competition, and differing baseline engagement rates.
One exception is Meta's Variance Reduction System (VRS), an algorithmic system that ensures ad delivery is similar to the distribution of eligible users on the platform~\cite{Bogen_2023}. 
After being sued by the Department of Housing and Urban Development for enabling skewed ad delivery that favored users by race or gender, Meta committed to the development of a special ad delivery flow that would counter skewed ad delivery~\cite{Isaac_2022}. 
The VRS was rolled out in 2024.
Meta VRS is not an advertiser-facing tool, but rather an algorithm that automatically adjusts skew with respect to age, gender, and race.
Advertisers are mandated to disclose if their advertising campaign falls under special categories (housing, credit, employment), in which case ad delivery will be mediated by the VRS. 
Advertisers can also choose to label their campaign as a special category if they want their ad delivery adjusted by the VRS. 
However, opting into VRS delivery does not meaningfully improve advertiser control or transparency, while significantly increasing costs~\cite{imana2025externalevaluation}.
There are several ways in which VRS does not improve advertiser control.
First, while the VRS is intended to specifically prevent discrimination toward legally protected groups, public interest advertisers may seek to measure and reduce skew along other axes of historical disadvantage, such as for Spanish language speakers (rather than focusing on ethnicity directly)~\cite{koenecke2023popular}. 
Second, advertisers may seek an audience different from the distribution of eligible users on the platform; for example, the advertiser may seek for ad delivery to reflect local demographics or community preferences (as elicited through participatory design)~\cite{koenecke2023popular}.
Third, as expected, the Meta VRS only applies to Meta platforms.

An alternative approach to the Meta VRS is a \textit{budget split intervention}.
Recently, budget split interventions have been proposed as a strategy to reduce skew~\cite{imana2025externalevaluation, koenecke2023popular}.
Under this budget split intervention, which we term the Direct Budget Split, advertisers replace a single campaign, which applied no demographic targeting, with multiple new campaigns, each targeting a demographic group directly.
For example, we could imagine a case on Google Ads where an advertiser runs a single advertising campaign with no targeting applied.
Upon evaluating the campaign for skew, the advertiser finds ads had been consistently under-delivered to users over 65.
The advertiser could then implement a Direct Budget Split by replacing their original campaign with two new campaigns, one targeting users under 65 and one targeting users over 65. 
The advertiser could allocate budget to each campaign such that overall ad delivery satisfies their definition of ``fair'' (e.g., proportional to local demographics, or in accordance with community preferences). \citet{imana2025externalevaluation} used a lab-designed case study to compare Meta VRS to the budget split intervention, and found that the budget split intervention categorically outperformed the Meta VRS by providing greater advertiser control at a lower cost.

We build from~\citet{imana2025externalevaluation} by proposing a novel design for a budget split intervention that includes users with unknown demographics, which we describe in Section 4.
We implement this intervention for a real-world advertising campaign run by a government agency, and evaluate the results of the intervention in Section 5.

\subsection{Trade-offs of Demographic Data for Advertising}
\label{sec:advertising tradeoffs}
Platform-inferred demographic labels introduce measurement error into estimates of ad delivery skew~\cite{imana2023having}. 
These demographic labels are drawn from users' self-selected settings or inferred through browsing behavior, and used by advertisers to implement demographic targeting or analyze engagement by audience segment~\cite{googleAboutDemographic}. 
Advertising platforms do not publicly share estimates of data accuracy, with external audits reporting conflicting degrees of accuracy by platform~\cite{tschantz2018accuracy, sabir2022analyzing, venkatadriAuditingOfflineData2019}. 
Alternatively, advertisers can purchase demographic data from data brokers, which gather information by purchasing traffic data from individual websites and collecting user activity on social networks~\cite{bergemann2019MarketsforInformation}.
However, purchasing data from data brokers might not improve accuracy, may incur ethical risks, and may increase advertising costs.
Prior work has found the accuracy of data brokers' demographic labels to be worse than a coin flip~\cite{neumann2019frontiers, venkatadriAuditingOfflineData2019}, while presenting ethical concerns through the ``commodification of personal data,'' which is fundamentally at odds with consumer power and privacy~\cite{craindatabrokers}.
As such, for government advertisers to participate in the data broker economy could undermine local trust and threaten individual privacy~\cite{collier2022ukadvertising}.

Researchers auditing ad delivery have circumvented the issues of demographic data collection with publicly available voter records~\cite{imana2021jobads, ali2019discrimination, aliAdDeliveryAlgorithms2021, sapiezynski2024proxies, imana2025inferreddemographics, sapiezynskiAlgorithmsThatDont2022, kaplan2022impliedidentityaddelivery}.
Methods to infer user attributes, such as Bayesian Improved Surname Geocoding (BISG), could also complement platform-provided demographic labels~\cite{imana2025inferreddemographics}.
However, these methods require personal user data.
For example, BISG requires individual surnames and residential addresses.
To infer users' demographics, advertisers would need to purchase data from data brokers, which has financial and ethical costs.
As such, government advertisers may be limited to platform-provided demographic labels.

By design, leading methods for ad delivery audits do not include users with missing demographic features. 
However, advertisers on Google Ads inevitably encounter ``unknown'' users, for whom Google does not infer a demographic label. 
Google's gender inferences are presented as three categories: male, female, and unknown; Google Ads allows advertisers to target male-inferred or female-inferred users.
While some social media platforms (e.g., Meta) have a high share of logged-in users and thereby more user data with which to infer user demographics, search platforms (such as Google) may have a higher share of logged-out users, making user identities harder to infer~\cite{tschantz2018accuracy}.
While advertisers could ignore unknown users, data missingness is non-random. 
Users without inferred labels are more likely to have low online footprints, which correlates with low socioeconomic status, or have non-binary gender identities~\cite{neumannDataDesertsBlack2024, Merrill_2021}.
To consider only Google's ``known'' gender labels (i.e., male, female) could reify larger systems of heteropatriarchal power, whereby transgender, non-binary, and gender non-conforming individuals are devalued and marginalized~\cite{bivens2019facebookgender, sampson2023queerpeoples, bivens2016baking}.
Excluding users with the Google-inferred unknown label may systematically exclude socioeconomically disadvantaged and non-binary individuals from outreach campaigns conducted through online advertising.

To use platform-inferred demographic labels requires grappling with competing harms: on one hand, online targeting denies users agency; on the other hand, without addressing ad delivery skew, some users would systematically receive less information about opportunities.
Platform-inferred gender labels hinder users' autonomy by denying their right to control their algorithmic identities, while reifying the essentializing project of defining gender through a hetero-normative binary (male, female)~\cite{barbosaWhoAmDesign2021, karizatAlgorithmicFolkTheories2021, shekhawat2019algorithmicprivacy, bivens2016baking}.
Improving users' control over demographic inferences would not eradicate concerns of user agency: users may never feel true agency in an advertising system that monetizes attention and incentivizes the collection and sale of personal data~\cite{hampton2021blackfeminist, sampson2023queerpeoples, zuboff2019surveillancecapitalism, Cannon2022nonbinary, craindatabrokers}. 

Despite the harm to user agency, platforms continue to infer user demographics from browsing history because it is profitable.
Granular targeting is desirable to advertisers that want to conduct strategic targeting and gain market insights into the users engaging with their advertisements. 
In addition to their desirability for commercial advertisers, inferred demographic labels enable the operationalization of fairness audits, such as evaluating ad delivery for skew.
The tension between fairness audits and user autonomy reflects broader fairness-privacy trade-offs inherent in disparity audits~\cite{king2023PrivacyBiasTradeoff}.
In our work, we develop a strategy for evaluating and addressing ad delivery skew with the imperfect demographic labels provided by Google.
Considering the inherent cost of online advertising to user self-determination, in Section 6 we make recommendations for government advertisers to also consider traditional, offline approaches for outreach, and call for online platforms to heed prior work demanding improved design for user agency.

\section{Case Study}
We conducted a case study to analyze ad delivery skew in collaboration with a state-level government agency that managed ad campaigns for an initiative supporting small business owners.
In particular, the campaigns were intended to raise engagement with a government-developed website with information, consultative services, and grants for entrepreneurs.
Historically, women had been underrepresented among business owners in the state, as is the case across the United States.
In 2023, the U.S. Census Bureau reported that only 39\% of 36.4 million U.S. businesses were owned by women~\cite{uscensus_business_owner_characteristics_2025}. 
The agency sought to prevent worsening gender disparities in business ownership by ensuring their campaigns reached users in proportion to state population demographics (50\% male, 50\% female).

In Section 3, we describe the original campaigns' set-up, using data collected over April-November 2024, and evaluate ad delivery skew by gender. 
In Section 4, we describe our budget split intervention design, which we ran from January-February 2025. 
In Section 5, we report the results of the budget split intervention. 

\subsection{Case Study: Advertising Terminology}
\label{subsection: case study advertising terminology}
Here, we introduce advertising terminology as defined for the context of our case study.
The advertising engagement pipeline has three core performance metrics: impressions, clicks, and conversions. 
Campaigns were run on multiple online advertising platforms, but we elect to study Google Search as it was the platform on which the agency spent the most money in the prior year.
Furthermore, Google Search is a dominant platform for paid search advertising: as of March 2025 Google Search was estimated to capture 90\% of search market share~\cite{murraygooglesearch}.
In paid search, advertisers bid on keywords, with winning ads appearing in the ``Sponsored'' section of Google search results.
Impressions refer to the count of ad views by Google users, and serve as a measure of ad delivery.
Clicks are defined as the count of times Google users click on the ad. 
Conversions are defined as actions taken within the website following a click.
In this case, conversions are second clicks on particular links within the website that indicate meaningful engagement with the website's public resources.
Advertisers also consider the Click-Through Rate (CTR) and Impression to Conversion Rate (henceforth CVR) as indicators of success. 
The CTR is the ratio of clicks to impressions. 
The CVR is the ratio of conversions to impressions. 
We report cost-effectiveness through the Cost Per Mille (CPM), or the cost per 1,000 impressions. 
When referring to cost, we consider dollars spent rather than budget: while advertisers set budgets for their campaigns, automated bidding may underspend or exceed daily budgets.
We attribute each of these metrics to user demographic groups using Google-inferred demographic labels. 
Google Ads charges per click, so spend is attributed to the demographic groups of the users whose clicks generated spend.

\begin{table}[h!]
\centering
\begin{tabular}{p{0.22\linewidth} p{0.74\linewidth}}
\textbf{Campaign Setting} & \textbf{Description} \\ \hline
Creatives &
Google Search ads consisted of a website headline and a description of the website.\\[0.4em]

Keywords &
Keywords were drawn from a list of business terms, often paired with the state name. \\[0.4em]

Budget &
Average daily budget was set at \$65 per campaign. \\[0.4em]

Targeting &
Geographic targeting was applied to limit delivery to the government agency's state. \\[0.4em]

Cost Per Action (CPA) & 
The target CPA (dollars spent per conversion) was set at \$2.50 on average. \\[0.4em] 

Clicks & 
Clicks were defined as clicks on the advertisement. \\[0.4em]

Conversions & 
Conversions were defined as second clicks that show engagement with website resources.\\[0.4em]

\end{tabular}
\caption{Original Campaign Parameters}
\label{tab:campaign setup}
\end{table}
\subsection{Original Campaign Set-Up}
The original campaign set-up was designed by a third-party marketing vendor to whom the government agency outsourced advertising operations.
The vendor designed a strategy to reach Google Search users with interest in the initiative through two advertising campaigns.
One campaign used the \textit{maximize clicks} bidding strategy, which automatically sets bids to maximize clicks within a daily budget.
The second campaign used a \textit{maximize conversions} bidding strategy, an automated bidding strategy to maximize conversions within a daily budget.
We refer to the campaign that maximized clicks as ``Max Clicks'' and the campaign that maximized conversions as ``Max Conversions.''
Campaign settings were identical other than their bidding strategy, as noted in Table~\ref{tab:campaign setup}.

\subsection{Definition of Skew}
\label{subsection: skew definition}
To evaluate skew in ad delivery, we draw from the delivery ratio used by the Meta VRS~\cite{imana2025externalevaluation}. 
Our definition of ad delivery skew for a given demographic attribute, whose groups \(g_i\) form the set \(G\), is as follows:
\(
\textbf{Skew}_{g',c} = \frac{\text{Impressions}_{g',c}}{\sum_{g_i \in G}\text{Impressions}_{g_i,c}} 
\).
Here, \(g'\) is the group that skew is measured with respect to (e.g., male), and \(c\) refers to the advertising campaign (e.g., Max Clicks).
\(\text{Impressions}_{g_i,c}\) corresponds to the count of impressions attributed to group \(g_i\) for advertising campaign \(c\). 
We report gender-based skew as the ratio of impressions by Google-inferred male users, relative to impressions by Google-inferred male and female users. 
This ad delivery skew metric can be expanded to other engagement metrics (e.g., spend, clicks, conversions) by replacing impressions with the relevant metric.
We exclude unknown users from the calculation of skew because our baseline, state demographic data, is in terms of male and female users. 
The government agency had a normative expectation of parity in delivery of ads to male and female users, but had no expectation regarding the prevalence of unknown or non-binary users. 
If advertisers were to use alternative approaches to set normative expectations for ad delivery, such as reducing variance from the distribution of the eligible audience, the above definition could be expanded to include unknown users.
We follow previous literature and consider ad delivery to reach parity if the 99\% confidence interval of ad delivery skew includes 0.5~\cite{imana2025inferreddemographics, imanaAuditingEducationAds2024, ali2019discrimination}. 
When measuring skew for male users, values over 0.5 would reflect skew toward male users, while values below 0.5 reflect skew toward female users. 

\subsection{Result: Consistent Skew in the Original Campaign}
\begin{figure}[ht]
  \centering
  \begin{subfigure}[b]{0.44\textwidth}
    \centering
    \includegraphics[width=\textwidth]{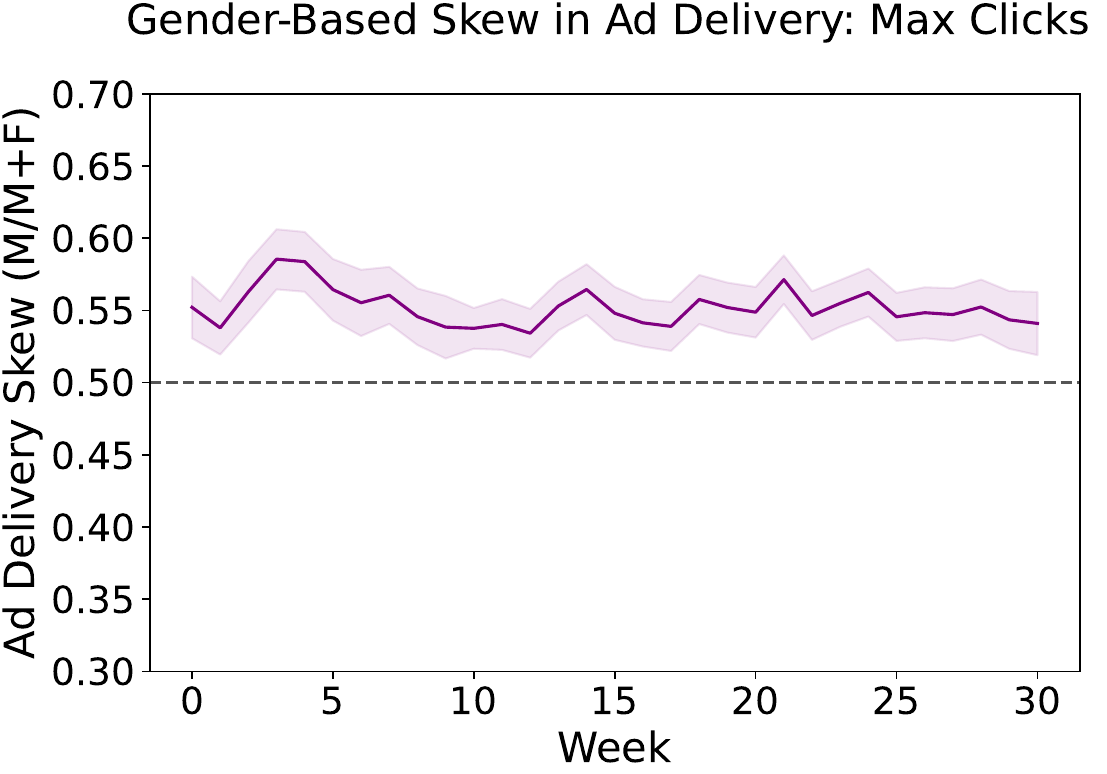}
    \caption{}
  \end{subfigure}
  \hfill
  \begin{subfigure}[b]{0.47\textwidth}
    \centering
    \includegraphics[width=\textwidth]{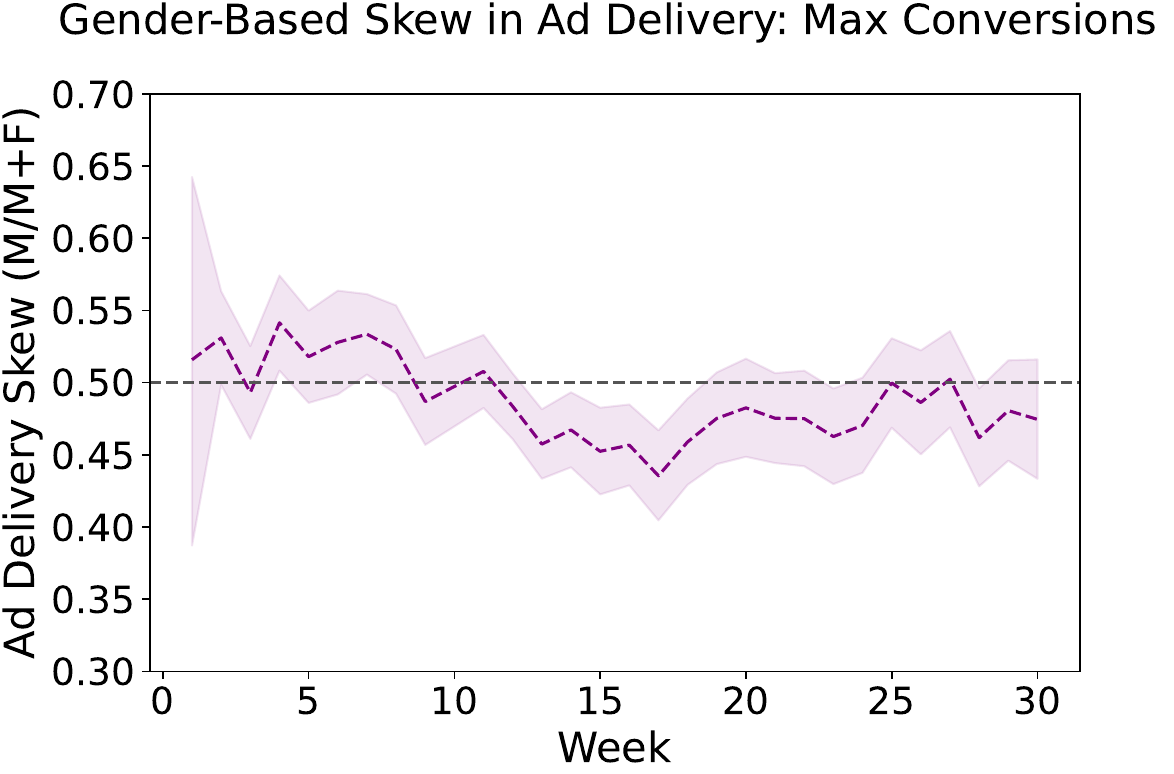}
    \caption{}
  \end{subfigure}
\caption{Skew in ad delivery over 31 weeks in (a) Max Clicks (solid line) and (b) Max Conversions (dashed line) campaigns. 
Ad delivery skewed toward male-labeled users over the 31-week period in the Max Clicks campaign while the Max Conversions campaign did not display consistent skew in either direction, despite female-labeled users having a higher conversion ratio. 
99\% CIs are shown.}
\Description{Two time series which show skew for the original campaigns, one maximizing clicks and one maximizing conversions. Skew in ad delivery over 31 weeks in (a) Max Clicks (solid line) and (b) Max Conversions (dashed line) campaigns. 
Ad delivery skewed toward male-labeled users over the 31-week period in the Max Clicks campaign while the Max Conversions campaign did not display consistent skew in either direction, despite female conversions having a higher conversion ratio. 
99\% CIs are shown.}
\label{fig:skew_over_time}
\end{figure}

\textbf{Original Max Clicks campaign ad delivery was consistently skewed toward users labeled male.}
We report ad delivery skew (y-axis) in each campaign over the 31 weeks (x-axis) of each original campaign.
Figure~\ref{fig:skew_over_time} shows that Max Clicks ad delivery consistently skewed toward male users, while skew in Max Conversions fluctuated and did not demonstrate consistent skew in either direction.
We cannot isolate the reasons for skew in the Max Clicks campaign due to the black-box nature of real-time bidding systems (see Appendix Figure 6a).
However, one hypothesis is a gender-based difference in baseline engagement rates: we observed in the Max Clicks campaign that the aggregate CTR for male-labeled users was higher than for female-labeled users.
Therefore, a campaign that seeks to maximize total clicks (i.e., Max Clicks) would benefit from increasing delivery to male-labeled users. 
In comparison, in the Max Conversions campaign, female-labeled users have marginally higher CVR than male users (see Appendix Figure~\ref{fig: cvr and ctr}b).
Therefore, a campaign seeking to maximize conversions (i.e., Max Conversions) would benefit from delivering more ads to female-labeled users, though we do not observe this in the historical data.
This engagement-based optimization may inherently drives skew towards male-labeled users in Max Clicks and female-labeled users in Max Conversions, which is counter to the agency's goal of equitable outreach.

\section{Method: Budget Split Intervention}
Budget splitting has been proposed as a cost-effective and interpretable approach to reducing skew in ad delivery~\cite{imana2025externalevaluation}.
As described in Section 2, budget splitting replaces a single original campaign with multiple new campaigns with group-specific targeting. 
By controlling the amount of budget attributed to each new campaign, the advertiser can adjust the share of ad delivery to each group.
We build on prior literature by designing and implementing a budget split intervention with unknown users in the context of government advertising.   

\subsection{Including Unknown Users for Government Outreach}
\label{subection: unknown users in budget split}
Google provides advertisers with three labels for user gender: male, female, and unknown. 
While previous research in auditing ad delivery skew has only considered users with known demographic attributes, there are several reasons government advertisers might prefer to include unknown users in their audience. 
First, excluding users is misaligned with the goal of ensuring anyone in the state could potentially be reached by ads raising awareness of public resources. 
Second, systematically excluding unknown users would be particularly problematic as prior research suggests these users are more likely to be individuals with low socio-economic status or non-binary gender identities~\cite{neumannDataDesertsBlack2024, bivens2019facebookgender, Merrill_2021}.
Third, since a significant portion of users on Google Search have unknown labels~\cite{tschantz2018accuracy}, excluding these users would diminish the size of the potential audience and could increase competition and costs if many advertisers are bidding for users with ``known'' demographic labels~\cite{Ahmadi2024overwhelmingTargeting}.
To avoid these concerns, we build on the Direct Budget Split proposed in prior literature and design a budget split with unknown users. 

\subsection{Budget Split with Unknown Users Design}
Google Ads does not enable advertisers to directly target unknown users as a standalone group. 
While the specific reason for this design choice is not publicly available, the Google Ads website states that \textit{``the ``Unknown'' demographic category is selected by default because you can reach a significantly wider audience''}~\cite{googleAboutDemographic}.
However, Google Ads allows advertisers to target unknown users together with known user groups (i.e., male, female labels), which we leverage to split our original campaign into four new campaigns with different target audiences by Google-inferred label: male, female, male and unknown, and female and unknown.

\begin{figure}[ht]
  \centering
  \includegraphics[width=.8\textwidth]{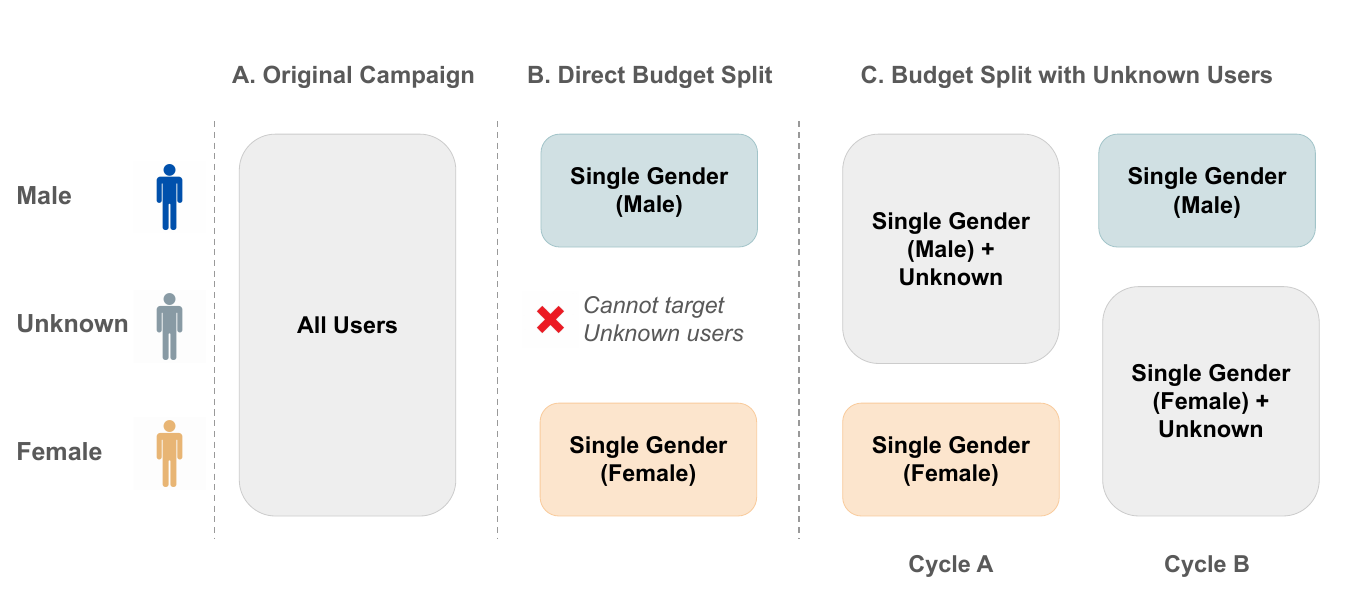}
  \caption{
  A. Original Campaign: One campaign, targeting all users (male, female, unknown). 
  B. Direct Budget Split: Budget split proposed in prior literature. Two campaigns run in tandem, each targeting a single group (male or female).
  C. Budget Split with Unknown Users: Our contribution of a novel budget split design. 
  Four campaigns run in two cycles (cycle A and cycle B). 
  Cycle A includes one Single Gender campaign (female) and one Single Gender + Unknown campaign (male + unknown).
  Cycle B includes one Single Gender campaign (male) and one Single Gender + Unknown campaign (female + unknown).
  }
  \Description{A graphic that shows the budget split intervention, as it compares to two other approaches for targeting: all-gender (the original campaign), or a direct budget split (as prior work has used).}
  \label{fig: budget split}
\end{figure}

As depicted in Figure~\ref{fig: budget split}, prior work has included the agency's ``Original Campaign'', which used ``All Users'' targeting; or the ``Direct Budget Split'' from~\citet{imana2025externalevaluation}, which used single-group (i.e., ``Single Gender'') targeting. 
In contrast, we propose a ``Budget Split with Unknown Users''; because unknown users cannot be targeted directly on Google Ads, we instead include them in ``Single Gender + Unknown'' campaigns by excluding the other binary gender label (i.e., male or female) from targeting. 
We then incorporated a combination of Single Gender and Single Gender + Unknown campaigns.
We avoid increasing competition by alternating between two cycles: cycle A with one campaign targeting female users and a second campaign targeting male and unknown users; cycle B with one campaign targeting male users and one campaign targeting female and unknown users. 
We alternated the day of the week and time of day for which each cycle was active over the course of the six-week period during which the budget split campaigns ran.

The budget for each new campaign can be calculated with the desired baseline ratio and group cost-effectiveness.
In the absence of priors, we initialized the budgets to reflect the desired skew ratio. 
With a desired ratio of 0.5, we allocated 50\% of the budget to male-targeting campaigns and 50\% of the budget to female-targeting campaigns. 
We applied this budget splitting design for each of the two initial campaigns (Max Clicks, Max Conversions).

\section{Results}
The Budget Split with Unknown Users serves as a useful middle ground between the ``Original Campaign'' and ``Direct Budget Split'' (see Figure~\ref{fig: budget split}). 
Including unknown users reduces skew relative to the Original Campaign and improves cost-effectiveness relative to Direct Budget Split.\footnote{Due to resource constraints, we did not run the Direct Budget Split as a standalone experiment, but rather approximated via Single Gender campaigns.
Further, our campaigns ran in cycles, unlike budget splits in prior work, to prevent increasing competition.
}
We provide empirical results on observed skew (Section 5.1), cost-effectiveness (Section 5.2), and simulated distributions of unknown users (Section 5.3). 

\subsection{Budget Split with Unknown Users Reduces Skew}
\begin{figure}[ht]
  \centering
  \begin{subfigure}[b]{0.45\textwidth}
    \centering
    \includegraphics[width=\textwidth]{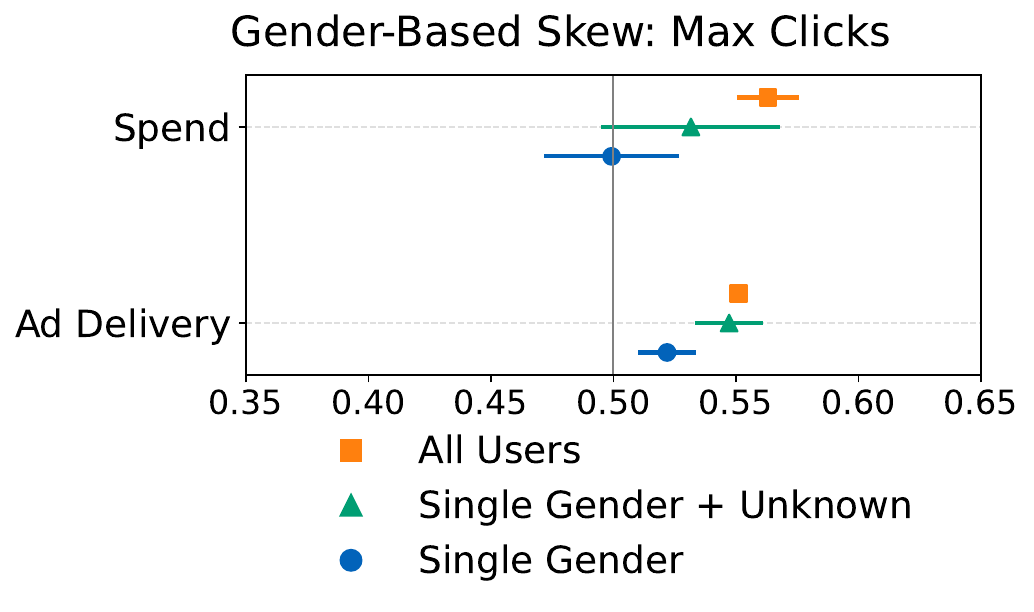}
    \caption{}
    \label{subfig: skew max clicks}
  \end{subfigure}
  \hfill
  \begin{subfigure}[b]{0.45\textwidth}
    \centering
    \includegraphics[width=\textwidth]{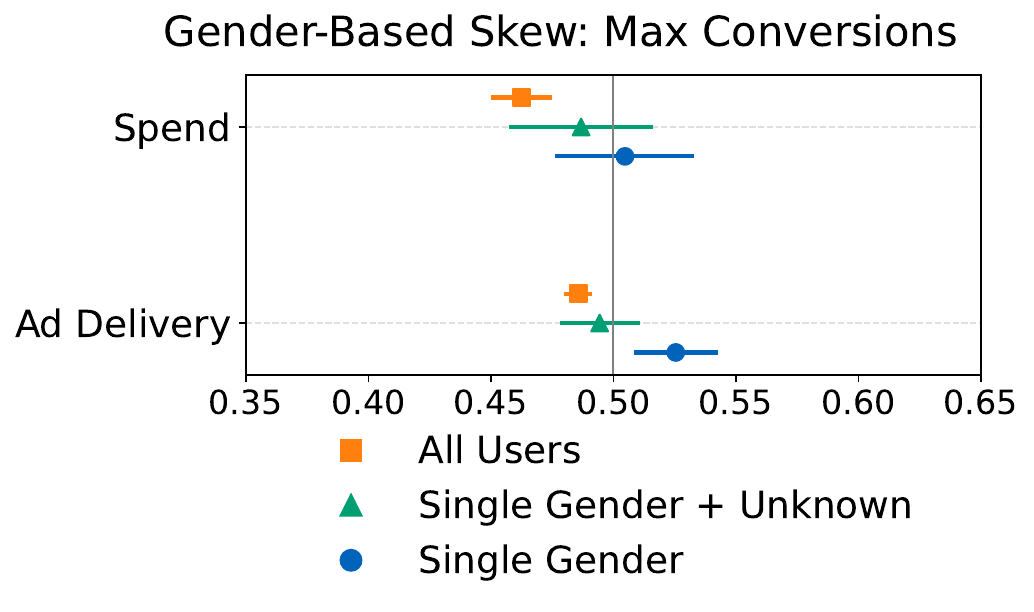}
    \caption{}
    \label{subfig: skew max conversions}
  \end{subfigure}
  \caption{
  Budget split intervention reduced skew in spend and delivery of (a) Max Clicks and (b) Max Conversions campaigns. 
  Each point reflects estimates of skew, as defined in Section 3.3.
  Figures show the share of male users for each level of targeting (All Users, Single Gender + Unknown, and Single Gender).
  Estimates for the All Users targeting schema reflect aggregate engagement over the 31 weeks of the Original Campaigns (Fig. 1).
  99\% CIs with an Agresti and Coull correction are shown.}
  \Description{Point-range plots that show estimates for skew with 99\% CI. The figure shows that skew is reduced under the budget split intervention, as the estimate (point) shifts toward the desired ratio (0.5).}
\label{fig:skew}
\end{figure}

In Figure~\ref{fig:skew} we show skew (x-axis) for spend and ad delivery (y-axis). 
Values to the left of the vertical 0.5 line indicate skew toward female-labeled users, to the right indicate skew toward male-labeled users.
We report skew in ad delivery with the formula described in Section~\ref{subsection: skew definition} as the share of spend attributed to male-labeled users, relative to the total spend attributed to users labeled male or female.
We compare skew between the Original Campaign and the Budget Split with Unknown Users.
In particular, we show skew by campaign targeting granularity level, as shown in Figure~\ref{fig: budget split}: All Users (Original Campaign), Single Gender + Unknown (Budget Split with Unknown Users), and Single Gender (Budget Split with Unknown Users).
Skew for All Users is calculated with aggregate spend and impressions counts. 
Skew for the Budget Split with Unknown Users campaigns is calculated with spend and impressions aggregated to the level of targeting granularity (i.e., Single Gender or Single Gender + Unknown). 
For example, the skew in spend reported for Max Clicks Single Gender (Figure~\ref{subfig: skew max clicks} dark blue circle, top row) is calculated with aggregate spend from the Single Gender campaigns (male, female) that were run in cycles A and B of the Budget Split with Unknown Users.

\textbf{Budget Split with Unknown Users enforces parity in spend.}
As shown in the top row, ``Spend'', of Figure~\ref{subfig: skew max clicks}, skew in spend in the original Max Clicks campaign (All Users) was greatest (skewing towards male users) at 0.563, calculated using the aggregate spend over the 31 weeks of the Original Campaign. 
Skew in spend was lowest in the Single Gender campaigns, achieving near-gender-parity at 0.499, calculated using the aggregate spend over the 6 weeks of the budget split Single Gender campaigns (i.e., male alone, female alone). 
In comparison, skew in spend for the Single Gender + Unknown campaigns fell between the two alternate targeting strategies, slightly skewing towards male users at 0.532 -- calculated using aggregate spend over the budget split Single Gender + Unknown campaigns (i.e., male and unknown, female and unknown). 
We note that the 99\% Confidence Interval for the Single Gender + Unknown campaigns ranges over [0.495, 0.568], which contains the desired gender-parity of 0.5.
In contrast, the All Users campaign 99\% CI [0.550, 0.576] does not contain 0.5, indicating that our intervention enforces parity in spend better than the Original Campaign (and comparably to, or slightly worse than, a Direct Budget Split intervention).
We observe consistent results in the Max Conversions Campaign (see Figure~\ref{subfig: skew max conversions}).
Our results confirm the findings of~\citet{imana2025externalevaluation} that budget splitting is a functional approach to increase advertiser control over the distribution of spend.
We extend prior work by showing that targeting unknown users effectively increases advertiser control over spend allocation while circumventing the concerns of excluding unknown users (as described in Section~\ref{subection: unknown users in budget split}). 

\textbf{Budget Split with Unknown Users shifts ad delivery skew toward the advertiser's desired ratio.}
In Max Clicks, ad delivery skew in the Original Campaign (All Users) was shifted from 0.550 (99\% CI=[0.548, 0.554]) to 0.522 (99\% CI=[0.510, 0.534]) under Single Gender targeting, and 0.547 (99\% CI=[0.533, 0.561]) under Single Gender + Unknown targeting.
Scaled to the Original Campaign's reach, a proportional reduction in skew would translate to delivering ads to about 400 more female-labeled users under Single Gender + Unknown targeting, and over 4,000 additional female-labeled users under Single Gender alone.
Single Gender targeting is expected to be more effective at reducing skew, as advertisers can control delivery to each group, while Single Gender + Unknown targeting introduces uncertainty, as advertisers cannot control how ad delivery is split between labeled-gender and unknown users within each campaign.
In Max Clicks, both budget split targeting strategies shifted skew in ad delivery toward the desired ratio of 0.5. 

In Max Conversions, ad delivery skew in the Original Campaign shifted from 0.486 (99\% CI=[0.480, 0.491]) to 0.526 (99\% CI=[0.508, 0.543]) under Single Gender targeting, and 0.494 (99\% CI=[0.478, 0.511]) under Single Gender + Unknown targeting, which is closest to the desired skew of 0.5.
Scaled to the Original Campaign's reach, a proportional shift in skew would translate to delivering ads to about 400 more male-labeled users under Single Gender + Unknown targeting, and over 2,000 additional male-labeled users under Single Gender alone.
Unlike under the Single Gender + Unknown approach, Single Gender targeting over-corrected by introducing ad delivery skew toward users labeled male.

We note that the larger confidence intervals, which were calculated using data aggregated at the daily level, for budget split estimates reflect the duration of the intervention: the Original Campaign ran for 31 weeks, while the budget split intervention ran for 6 weeks. 
The budget split duration was driven by financial constraints. We encourage future work to experiment with longer interventions.

The Budget Split with Unknown Users reduced skew in Max Clicks, serving as a middle ground between the Direct Budget Split, in which unknown users are excluded, and the Original Campaign, in which all users are targeted.
In the Max Conversions campaign, the Budget Split with Unknown Users similarly reduced skew from the Original Campaign; in contrast, the Single Gender campaigns over-corrected, shifting ad delivery skew toward users labeled male.
Due to the black-box nature of Google Ads, we cannot isolate the cause of the shifts.
However, we hypothesize the persistent skew in ad delivery toward male-labeled users under Single Gender targeting (for both Max Clicks and Max Conversions) may be attributable to the cost-effectiveness of granular targeting. 

\subsection{Budget Split with Unknown Users Improves Cost-Effectiveness}

\begin{figure}[ht]
  \centering
  \begin{subfigure}[b]{0.45\textwidth}
    \centering
    \includegraphics[width=\textwidth]{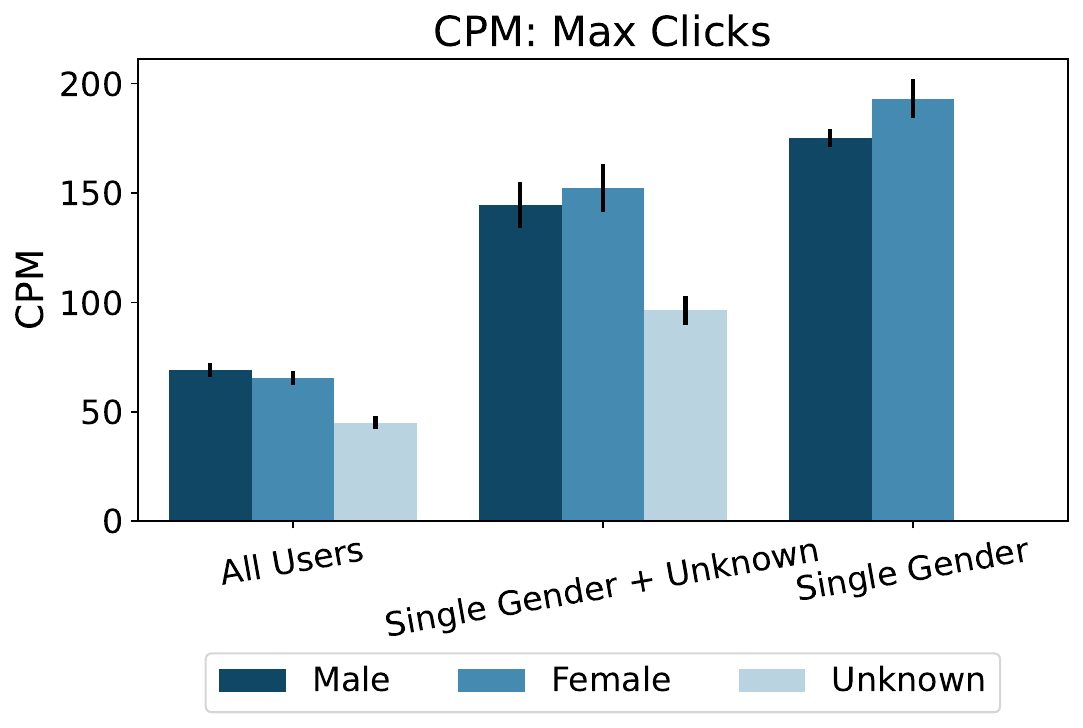}
    \caption{}
    \label{subfig: cpm max clicks}
  \end{subfigure}
  \begin{subfigure}[b]{0.45\textwidth}
    \centering
    \includegraphics[width=\textwidth]{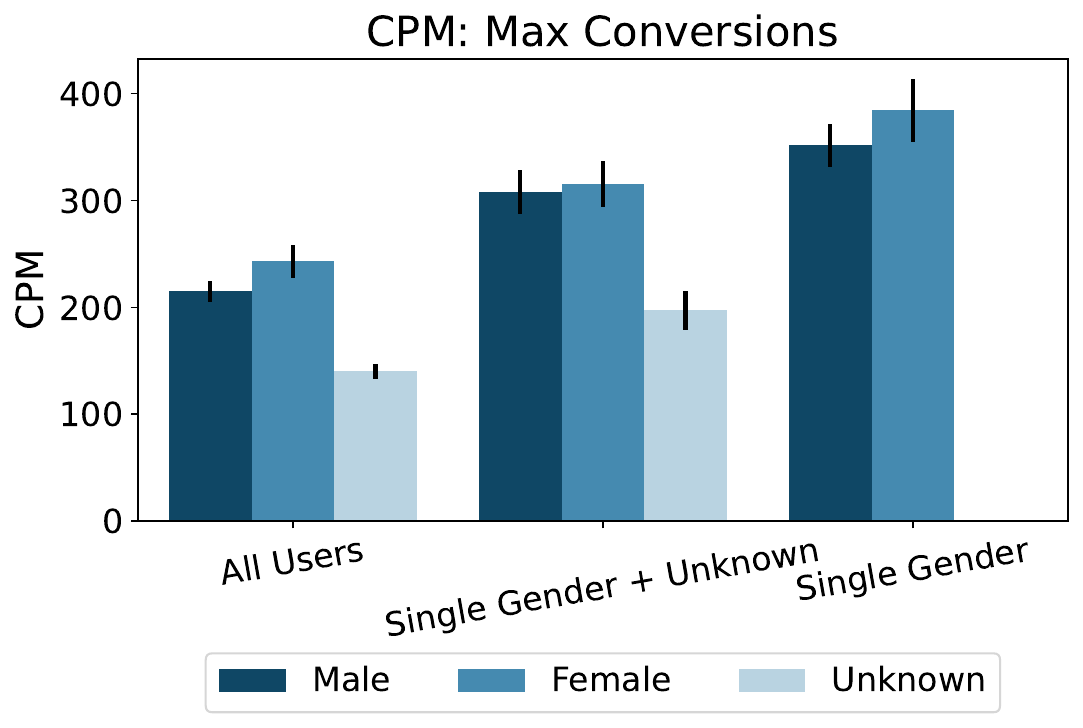}
    \caption{}
    \label{subfig: cpm max conversions}
  \end{subfigure}
  \caption{
  Average weekly CPM by Google-inferred gender label (i.e., male, female, unknown) and targeting approach in (a) Max Clicks and (b) Max Conversions campaigns. 
  Unknown CPM in Single Gender + Unknown category uses aggregated engagement and spend (attributed to unknown users) from Single Gender (male) + Unknown and Single Gender (female) + Unknown campaigns.
  Single Gender + Unknown targeting serves as a useful middle ground in cost-effectiveness between All Users and Single Gender targeting.
  Under Single Gender + Unknown targeting, the cost gap between female and male-labeled users is diminished, relative to the gap under Single Gender targeting. 
  Standard error bars are shown.}
  \Description{A barplot showing the cost per 1,000 impressions (CPM) for the three targeting approaches used in our budget split intervention. Costs increase (e.g. the height of the bars increase) with more granular targeting.}
  \label{fig:cpm}
\end{figure}

Cost-effectiveness is shown in Figure~\ref{fig:cpm}, in which each bar reflects the average weekly CPM for reaching users labeled male (dark blue), female (medium blue), and unknown (light blue) under each targeting strategy (x-axis).
No unknown users were reached in Single Gender campaigns (by design), so unknown CPM was not reported for that strategy.
As described in Section~\ref{subsection: case study advertising terminology}, CPM is calculated as spend divided by the count of impressions (and multiplied by 1,000).
For example, in Figure~\ref{subfig: cpm max clicks}, the Single Gender + Unknown male bar (dark blue) shows the average weekly CPM attributed to reaching users with the Google-inferred male label in the Single Gender + Unknown campaigns; this corresponds to the Single Gender (male) + Unknown campaign in cycle A of Figure~\ref{fig: budget split}.

\textbf{The Single Gender + Unknown targeting strategy reduces the difference in cost-effectiveness between users labeled male and female relative to Single Gender targeting.}
The persistent skew toward male-labeled users for the Max Clicks campaign shown in Figure~\ref{subfig: skew max clicks} might be explained by the gap in cost-effectiveness by gender: female-labeled users are more expensive than male-labeled users under Single Gender targeting, thus hindering the mitigation of ad delivery skew.
In Figure~\ref{fig:cpm}, for both Max Clicks and Max Conversions campaigns, under Single Gender targeting, the CPM of female-labeled users (medium blue) exceeds the CPM of male-labeled users (dark blue). 
This gap means that campaigns with the Single Gender targeting strategy paid more for impressions from female-labeled users than male-labeled users.
Consequently, equal budgets would result in greater delivery to male-labeled users.
This finding is consistent with prior work showing female users to cost more than male users, as described in Section~\ref{subsection: auditing skewed ad delivery}~\cite{ali2019discrimination,lambrecht2019algorithmic}.
However, when users were targeted in the Single Gender + Unknown condition, the gap in cost-effectiveness by gender was reduced substantially.
This finding suggests that a Budget Split with Unknown Users intervention can decrease the cost differential between male-labeled and female-labeled users, thus supporting equitable ad delivery, as a cost premium for female-labeled users could enable skew toward male-labeled users.
Our findings demonstrate the practical necessity of continuously updating budget allocations for budget split interventions in response to targeting cost premiums.

\textbf{The Single Gender + Unknown targeting strategy reduces the overall cost of budget splitting.}
We observe that the CPM increased with increased targeting granularity, indicating reduced cost-effectiveness.
While prior work has proposed budget splitting as more cost-effective than the Meta VRS system~\cite{imana2025externalevaluation}, we find that budget splitting still substantially increases advertising costs, relative to the All Users campaign. 
Indeed, the Direct Budget Split that has been proposed in prior work is the most expensive approach, as demonstrated by the Single Gender targeting strategy.
However, we find that the Single Gender + Unknown targeting strategy provides a middle ground in which advertisers can reduce skew, relative to the All Users condition, without incurring the costs of the Single Gender condition. 
As such, the Budget Split with Unknown Users reduces the cost trade-off for advertisers seeking to reduce skew in ad delivery. 

\subsection{Simulations of Unknown User Distribution Support Skew Estimates}
\begin{figure}[ht]
  \centering
  \includegraphics[width=.6\textwidth]{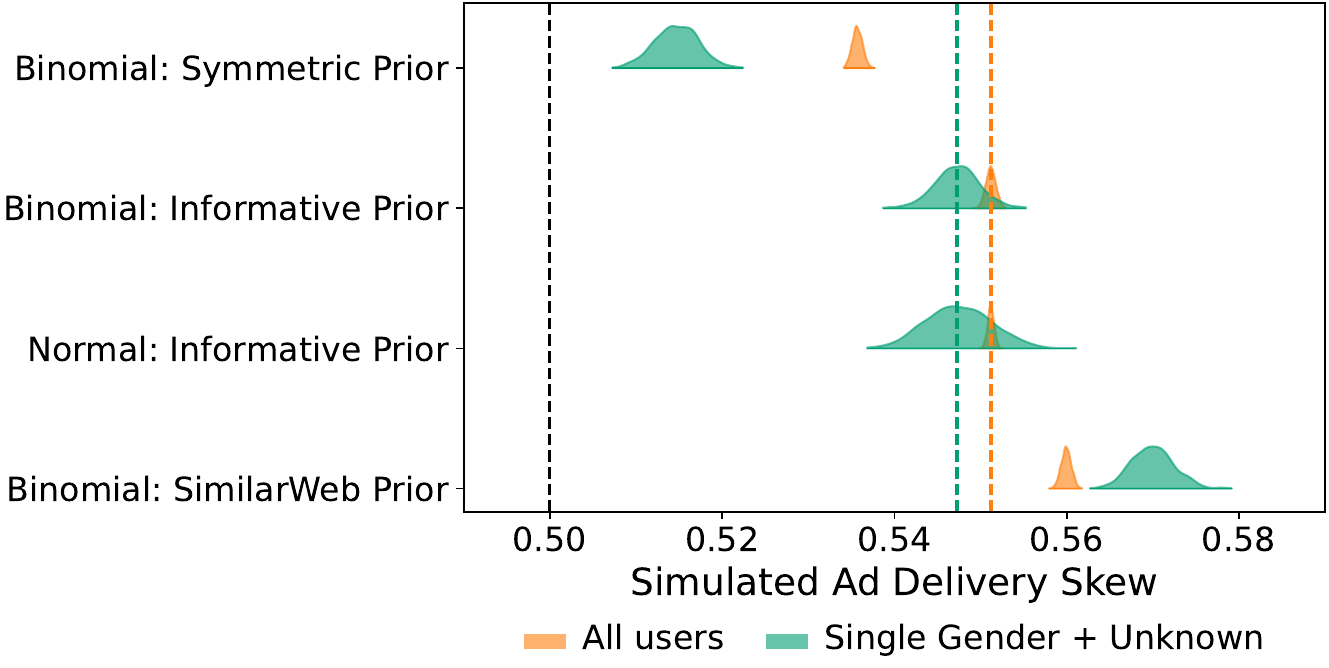}
\caption{Distributions of ad delivery skew for Max Clicks campaign under four sampling distributions (N=1,000). 
Vertical lines depict the estimated skew calculated with only Google-labeled male and female users for All Users (orange), Single Gender + Unknown Users (green), and desired skew ratio of 0.5 (black).} 
\Description{Each row shows the distribution of skew using simulated unknown user populations.}
\label{fig:max clicks unknown simulations}
\end{figure}

\textbf{Modeling skew under simulated gender distributions of unknown users reveals, under most models, lower skew for Budget Splitting with Unknown Users than the Original Campaign.}
Google's unknown demographic category likely reflects users that have greater predictive uncertainty than those users who received inferred labels~\cite{neumannDataDesertsBlack2024}. 
In order to understand if our estimates of skew toward male users would be changed dramatically with full knowledge of unknown users' gender identities, we estimate skew under different simulated distributions of unknown user gender.
We quantify the distribution of skew under uncertainty arising from unknown users (i.e., using reasonable priors and operating within Google's gender binary, we simulate how many of the unknown users are female or male).
Figure~\ref{fig:max clicks unknown simulations} shows the distribution of skew under 1,000 draws from four different simulated distributions of unknown users for the Max Clicks campaign. 
Under most models, distributions from the Single Gender + Unknown budget split intervention (green) have a modal point closer to the desired skew of 0.5 than the All Users original campaign (orange).
Expectedly, unknown user distributions that estimate a greater share of male users show increased skew toward male users, relative to our earlier estimates that calculate skew without unknown users (Figures~\ref{fig:skew_over_time},~\ref{fig:skew}).
We show consistent results for the Max Conversions campaign, in Appendix Figure~\ref{fig: max conversions simulations}. 

Our simulations draw from three settings (see details in Appendix Section~\ref{sec: simulation parameters}): (1) Symmetric Prior, where male and female users are equally likely to occur, (2) Informative Prior, where unknown users match the distribution of observed impressions from users labeled male and female, and (3) SimilarWeb Prior, drawn from the SimilarWeb marketing platform (estimated 58\% male users).
Each skew estimate is calculated by combining the observed count of users labeled male and female with the count of male and female users from the simulated unknown user population.  
We use these parameters for simulation as each poses a reasonable estimate for the distribution that unknown users may be drawn from.
We estimate the distribution of skew for All Users and Single Gender + Unknown campaigns, as the Single Gender targeting strategy does not reach any unknown users by design.
Under the Symmetric Prior and Informative Prior, we find estimates of skew are largely consistent with our estimates for skew using only male and female labeled users (Figure~\ref{fig:skew}).
Our simulations show increased skew when the unknown pool is heavily male-dominated (e.g., SimilarWeb Prior).
These simulations suggest that despite the uncertainty introduced with unknown users, our earlier evaluation of ad delivery skew in the presence of unknown users can still provide directional guidance for advertisers.

\section{Discussion}
Our case study confirms prior work that pernicious ad delivery skew can affect high-impact advertising domains, regardless of advertiser intent.
Advertisers have little recourse, as platforms do not offer satisfactory solutions~\cite{imana2025externalevaluation}. 
Instead, advertisers have to work within the constraints of platform-provided tools.
Budget splitting is an imperfect solution due to the fairness-utility trade-off from direct targeting, and the fairness-privacy trade-off from demographic inferences.
However, through the Budget Split with Unknown Users, we identify an opportunity to reduce skew while ameliorating the cost increase.
As such, we recommend that advertisers who have identified skew in ad delivery consider the Budget Split with Unknown Users to gain control over spend and ad delivery allocation, while mitigating the cost premium of granular targeting.
In addition, we provide recommendations that encapsulate lessons learned from our case study.
Finally, because skew can be influenced by facially neutral campaign settings, in Appendix Section~\ref{sec: checklist} we provide a practical checklist for advertisers. 

\subsection{Recommendations for Government Advertisers}
\subsubsection{Establish Technical Ownership During Public Procurement.}
Public procurement is how government agencies select private vendors from which to purchase goods and services.
It also serves as an opportunity for agencies to ensure that vendors can deliver the technical support required for program monitoring and skew evaluations.
In the context of AI procurement, \citet{johnson2025legacyprocurement} suggested that agencies build internal capacity for bias audits instead of relying on vendors, who may have misaligned incentives or lack domain expertise.
However, agencies may lack the budget or political buy-in to increase internal capacity~\cite{levin2010contracting, corpus2025}. 
When agencies outsource advertising operations, we recommend they use the procurement process to enforce expectations of a transparent and interoperable technical foundation. 

Agencies can use procurement to establish expectations for data collection, access, and storage.
In our case study, the vendor owned ad engagement data and did not maintain documentation.
Data was shared through ad-hoc support requests, which made data access slow and difficult. 
Our collaborator agency learned from this experience that their lack of data ownership blocked evaluations and audits of ad delivery skew.
The agency has now developed a data schema that it will use to request and store data from future vendors. 
While often taken for granted, strong expectations of data collection, access, and storage from vendors are critical to enable program monitoring and bias audits.

\subsubsection{Consider Offline Outreach to Historically Subordinated Groups}
As demonstrated by our case study, online advertising platforms present little recourse for inequitable ad delivery.
Interventions for reducing skew, such as budget splitting, increase costs and reduce total reach. 
Advertisers should consider if the value of online advertising outweighs the costs of fairness and inherent limitations of user agency posed by online targeting. 

While online advertising has the potential to reach a wide variety of people, it may not always be the right approach for public sector outreach. 
One example of unsuccessful online advertising outreach was a Georgia Department of Community Health (DCH) campaign, run by Deloitte, to raise enrollment for the Pathways to Coverage health insurance program.
Ultimately, the marketing campaign failed to reach target enrollment numbers, despite investing hundreds of thousands of dollars in Google and Meta advertising~\cite{coker2025firm}. 
External auditors recommended that the Georgia DCH instead consider alternative outreach, such as partnering with public libraries, faith-based organizations, and focus groups with under-enrolled communities such as people of color and rural populations~\cite{pathwaysreport}. 

\subsubsection{Harness Purchasing Power}
Large advertisers receive privileged treatment from online advertising platforms.
On Google, advertisers that reach certain thresholds of spend and engagement receive ``access to a wider range of benefits''~\cite{googleads_become_partner}.
While a single local government agency may not spend enough to reach partner status, if multiple agencies collaborated, they could unlock a better-supported advertising experience. 
Considering that government agencies may be unable to build in-house capacity, strengthening their access to external support without increasing any one agency's spend could prove beneficial to evaluating and addressing ad delivery skew. 

\subsection{Recommendations for Platforms}

\subsubsection{Provide Tools for Navigating Skew}
At a basic level, platforms could provide explanations for why ad delivery skew emerges and in-platform tools (e.g., dashboards) to measure and address skew by attribute.
Platforms may be hesitant to enable demographic budget splitting along sensitive attributes due to the potential for misuse. 
Instead, platforms can help raise awareness around ad delivery skew with informational resources about budget split interventions for domains and contexts that are not strictly regulated.

\subsubsection{Remove Bidding Constraints from Public Interest Grants}
Google Ads offers grants to nonprofit organizations. 
However, these grants mandate a maximum Cost Per Click (CPC) of \$2.00~\cite{google_ad_grants_details}.
This constraint may appear neutral, but can introduce skew against more expensive demographic groups, which has been shown to coincide with female-labeled users~\cite{lambrecht2019algorithmic}.
Platforms should remove CPC ceilings to avoid exacerbating skew.

\subsubsection{Make Transparent Demographic Inferences}
Some advertising platforms (e.g., LinkedIn, Amazon) do not make missing labels explicit.
However, an explicit ``Unknown'' category can be beneficial to avoid overstating confidence in inferences and enable varied budget split designs.
The gender labels Google provides advertisers (i.e., male, female, unknown) erase non-binary individuals.
However, adding gender identity categories would only provide an illusion of authenticity while maintaining the logic of data broker marketplaces and an essentialist definition of gender as discretizable~\cite{bivens2019facebookgender, Cannon2022nonbinary}.
Furthermore, to provide granular gender labels would threaten user privacy~\cite{korolova2010privacyviolations}, enable tokenization ~\cite{sampson2023queerpeoples}, and increase the vulnerability of non-binary users to algorithmic surveillance~\cite{Cannon2022nonbinary, hampton2021blackfeminist}.
\citet{bivens2016baking} suggest platforms instead share inferred demographic identities with users, with an opportunity for recourse and adjustment. 
While Google Ads already allows users to see their inferred demographic labels, future work could include usability studies to ensure recourse options are legible.
We note that transparency does not end harm experienced by marginalized individuals navigating the surveillance enterprise of online advertising~\cite{hampton2021blackfeminist, craindatabrokers, sampson2023queerpeoples}, but could potentially serve as an incremental improvement.

\subsection{Recommendations for Researchers}

\subsubsection{Limitations of Experimentation in the Wild}
Researchers should anticipate working within the constraints faced by their government collaborators, who are doing their best --- and incredibly impactful --- work with the constraints imposed upon them. 
These constraints may be financial, due to budget limits, or strategic, reflecting a commitment to program effectiveness.
Government collaborators may also face pressures about talking about fairness, especially with regard to sensitive attributes like race or gender. 
We provide a checklist in Appendix Section~\ref{sec: checklist} with guiding questions to consider when addressing skew in real-world advertising campaigns.

\subsection{Limitations}
Our findings demonstrate a principle from marketing research --- that more granular targeting can incur higher costs ---  in an algorithmic fairness context (i.e., addressing undesirable ad delivery skew). 
As such, the Budget Split with Unknown Users intervention can be extended to other domains, attributes, and platforms, with some limitations.
In particular, direct demographic targeting is prohibited in strictly regulated domains (e.g., housing, employment). 
Another limitation of our work is that our budget split intervention only applies for platform-inferred demographics, which are often sensitive characteristics.
For example, platform-provided attributes on Google Search include demographic features (e.g., age, language, gender, and household income), along with less sensitive affinity segments that reflect inferred user interests~\cite{google_audience_segments}.
Future work that tests our intervention in new domains, attributes, and platforms would be beneficial to explore the generalizability of our results.
Our intervention would be particularly relevant for other advertising platforms with high rates of unknown users, as is often the case for search. 
Also, while our results are consistent for the duration of our intervention, future work could confirm the magnitude of our results with a longer-running study.

\section{Conclusion}
Evaluating and reducing ad delivery skew is an important, albeit challenging, endeavor for government advertisers, in part because auditing methodology hasn't translated directly to advertising practice.
We summarize key takeaways from our research that build on auditing ad delivery skew: 
\begin{enumerate}
  \renewcommand{\labelenumi}{(\arabic{enumi})}
  \item \textit{In an ad campaign raising awareness of government resources, when the campaign maximized clicks, ads were consistently under-delivered to female-labeled users.}
  This finding demonstrates the importance of evaluating consequential ad delivery campaigns for skew by demographic attributes.
  \item \textit{We presented a Budget Split with Unknown Users intervention that avoids the costs of dropping users with unknown demographic labels.}
  This design considers the value of unknown users, as government advertisers may seek to include all users in outreach, regardless of Google's ability to infer demographic labels. 
  \item \textit{We demonstrate that the Budget Split with Unknown Users serves as an effective middle ground for advertisers, reducing costs relative to the previously proposed Direct Budget Split and reducing ad delivery skew relative to the Original Campaign.}
  Considering the steep costs of counteracting ad delivery skew, we propose an additional set of recommendations to government advertisers, online platforms, and researchers. 
\end{enumerate}

Our study raises the question: can online advertising systems, which are designed for for-profit advertisers, serve the public interest goals of government advertisers? 
Our research extends prior work to show that an unknown-aware budget split can increase advertisers' control over ad delivery, while ameliorating the financial trade-offs of budget splitting.
We encourage the FAccT community to ensure audit methodology can be adapted for real-world advertising, including consideration of unknown users.

\section{Endmatter}

\subsection{Generative AI Usage}
In the preparation of this manuscript, Claude Opus 4.5 was used to identify grammar errors. 

\subsection{Acknowledgements}
We thank Danaé Metaxa and Aleksandra Korolova for their thoughtful feedback on the project. We are also grateful to the participants of CODE@MIT 2025 and the Cornell Artificial Intelligence, Policy, and Practice (AIPP) initiative for helpful discussions of this work. 

\bibliographystyle{ACM-Reference-Format}
\bibliography{ref}

\appendix
\section{Supplementary Information}
\label{sec:si}
\subsection{Models for Simulation of Unknown Users}
\label{sec: simulation parameters}
In November 2025, SimilarWeb, a platform that aggregates browsing data and data broker reports, reported Google users to be 58\% male and 42\% female~\cite{similarweb_google_com}. 
Under the SimilarWeb Prior, we simulate the count of male impressions in the unknown user pool, \( U_M \), from \(U_M \sim Binomial(0.58, N_U)\), where \(N_U\) is the count of unknown user impressions.

While SimilarWeb estimates that most Google users are male, researchers have estimated more balanced use of the platform. 
For example, a survey conducted with Swiss participants found no significant difference in Google use between male and female users~\cite{festic2021long}. 
We simulate the Binomial: Symmetric Prior distribution with \( U_M \sim Binomial(0.5, N_U)\). 

The Binomial: Informative Prior is modeled with \(p=N_M/(N_M + N_F)\), \( U_M \sim Binomial(p, N_U)\) where \(N_M\) is the observed count of male impressions and \(N_F\) is the observed count of female impressions. 
Similarly, under the Normal: Informative Prior, \( U_M \sim \mathcal{N}(p, \sigma_p)\), with values truncated to the \([0,1]\) interval.
For the Max Clicks All Users simulation (Figure~\ref{fig:max clicks unknown simulations}, orange), we observed 55.12\% male users; for Max Clicks Single Gender + Unknown (Figure~\ref{fig:max clicks unknown simulations}, green) we observed 54.72\% male users; for Max Conversions All Users (Figure~\ref{fig: max conversions simulations}, orange) we observed 48.57\% male users; for Max Conversions Single Gender + Unknown (Figure~\ref{fig: max conversions simulations}, green) we observed 49.44\% male users.

An alternative approach to constructing the Symmetric and SimilarWeb priors would be to solve for the share of male users in the unknown user pool, such that the share of impressions attributed to male-labeled users of total impressions is centered at 0.5 (Symmetric) or 0.58 (SimilarWeb).
However, this would require an assumption that the distribution of impressions we observe is representative of eligible users on the platform. 
In Figure~\ref{fig:max clicks unknown simulations} we simulate this possibility under the Informative Prior; however, for the Symmetric and SimilarWeb priors we do not. 
This is because we anticipate that the distribution of male and female impressions we observed is not necessarily indicative of the distribution of eligible users on the platform, but rather potentially biased predictions of relevance, competition, or differing baseline engagement rates~\cite{ali2019discrimination, imana2025externalevaluation, imanaAuditingEducationAds2024,imana2021jobads, imana2023having}.
In Figure~\ref{fig: max clicks simulations with solve} we show simulations of the Symmetric and SimilarWeb Priors with Solve. 
In the case of the SimilarWeb Prior with Solve, the count of unknown users is simulated with the distribution \(U_M \sim Binomial(p_{SW}, N_U)\), where we solve for \(p_{SW}\) using \(0.58 = (N_M + U_M)/N_{Total}\).
Similarly, in the case of the Symmetric Prior with Solve, the count of unknown users is simulated with the distribution \(U_M \sim Binomial(p_{S}, N_U)\), where we solve for \(p_S\) using \(0.5 = (N_M + U_M)/N_{Total}\).
We observe that in these settings, skew estimates reflect the simulation parameters (e.g., 0.5, 0.58). 
We would only expect such simulations to be reasonable if observed ad delivery skew exactly reflected the true distribution of male and female labeled users on the platform.

\subsection{Advertising Strategy Checklist for Skew}
\label{sec: checklist}
\paragraph{Advertising strategy}
When establishing advertising strategy, there are many elements which appear neutral but which can influence the presence of skew. 
For example, \textit{platform choice} will be influential, as different platforms have different distributions of users. 
It is worth considering the limitations of choosing to advertise on a platform that may be dominated by a single group. 
As we showed in our case study, the \textit{bidding strategy} that drives automated bidding will influence the direction and magnitude of skew if there are differences in groups' baseline engagement rates across the advertising funnel (e.g., between clicks and conversions). 
As there is no theoretical grounding to suggest if users of a given demographic group will have different baseline engagement rates, this may only be solvable through testing and documenting skew under different bidding strategies. 
Maximum CPA (cost per action) is influential to skew because some groups of users are more expensive to reach than others, due to facing higher competition from other advertisers~\cite{ali2019discrimination}.
A low maximum CPA will prevent reaching expensive users, which may correlate with demographic groups.

\paragraph{Skew definition}
When defining skew, advertisers must ground their ideal value for skew with some baseline ratio.
Skew will then be evaluated as deviation from that baseline ratio.
There are many demographic attributes that could be considered when seeking to evaluate skew; however, advertisers may not have the budget or capacity to test for all. 
Therefore, advertisers should carefully consider \textit{attribute selection}.
One point of consideration should be which groups have been historically systematically excluded from the opportunities that are being advertised, so as to minimize harm to the most disadvantaged groups. 
Another consideration is data access. 
If an advertiser has low budget and capacity, they might be constrained to in-platform demographic features. 
If they have a higher budget, they might purchase from data brokers.
If they have higher capacity and budget, they might purchase data and infer user attributes as in~\citet{imana2025inferreddemographics}.
Advertisers must also consider \textit{threshold selection} to define a baseline ratio against which skew will be evaluated.
This baseline might be set with population demographics, as in our case study, participatory design~\cite{koenecke2023popular}, or eligible users on the platform~\cite{imana2025externalevaluation}, depending on the normative expectations of the advertiser. 

\paragraph{Budget split design}
When running a budget split, advertisers will need to decide whether or not to include unknown users. 
As we demonstrate in our case study, including unknown users can ameliorate the cost premium of granular targeting under the budget split, but reduce control over skew. 
We recommend advertisers consider including unknown users, to prevent users for whom the platform struggles to make a demographic inference from losing the opportunity to gain information on public services. 
As the budget split requires the development of new campaigns, advertisers will need to account for the \textit{cold start} period, in which the ad delivery platform is learning to optimize the campaign.
Engagement metrics during this period are likely to be unstable. 
The cold start period will differ by context and campaign; use historical data to estimate the duration and budget for any budget split tests to have longer durations. 
Finally, the advertiser should expect to \textit{iteratively split budget} attributed to each sub-campaign. 
As our case study demonstrated, the relative cost of reaching users may change under different targeting approaches.
Therefore, the amount of budget portioned into each arm of the campaign will need to vary by the cost-effectiveness of each group to achieve the desired threshold.

\begin{table}[ht]
\centering
\caption{Checklist of Factors Influencing Ad Delivery Skew}
\label{tab:checklist_skew}
\begin{tabular}{p{0.20\textwidth} p{0.20\textwidth} p{0.52\textwidth}}
\textbf{Factor} & \textbf{Stage of Evaluation} & \textbf{Guiding Questions} \\
\hline
Platform choice & Advertising strategy &
How does the eligible audience differ across platforms? \\
Bidding strategy & Advertising strategy &
How do baseline engagement differences across groups interact with the chosen bidding strategy? \\
Maximum Cost per Action (CPA) & Advertising strategy &
How does skew change with CPA? \\
Attribute selection & Skew definition &
Which user attribute is at risk of disparate impact or quality-of-service harm? 
How much budget and capacity is available to invest in accessing demographic data? \\
Threshold selection & Skew definition  &
Which engagement outcome (e.g., impressions, conversions) matters most across the ad engagement funnel?
What is a reasonable theoretical grounding for a target threshold?\\
Handling unknown users & Budget split design &
Should unknown users be included, grouped with known users, or excluded, given platform targeting constraints? \\
Cold start & Budget split design &
Has the campaign run long enough to exit the learning phase and yield stable skew estimates? \\
Iteratively split budget & Budget split design &
How will finer-grained targeting affect cost, cost-effectiveness, and group-level delivery? \\
\end{tabular}
\end{table}

\begin{figure}[ht]
  \centering
  \begin{subfigure}[b]{0.4\textwidth}
    \centering
    \includegraphics[width=\textwidth]{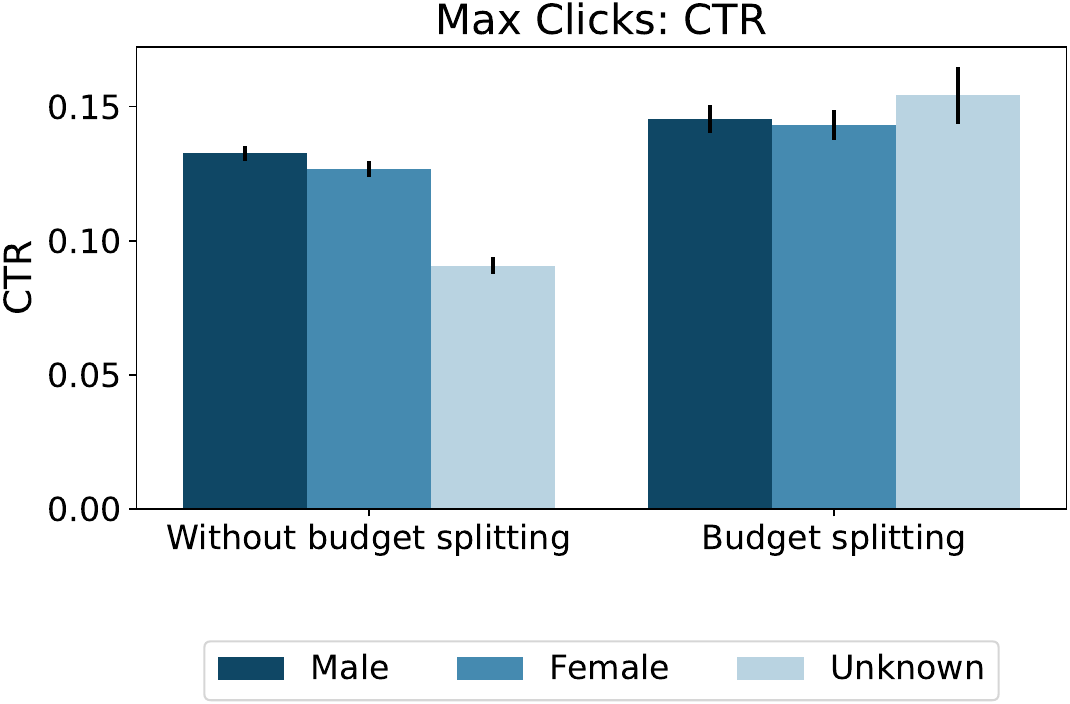}
    \caption{}
  \end{subfigure}
  \hfill
  \begin{subfigure}[b]{0.4\textwidth}
    \centering
    \includegraphics[width=\textwidth]{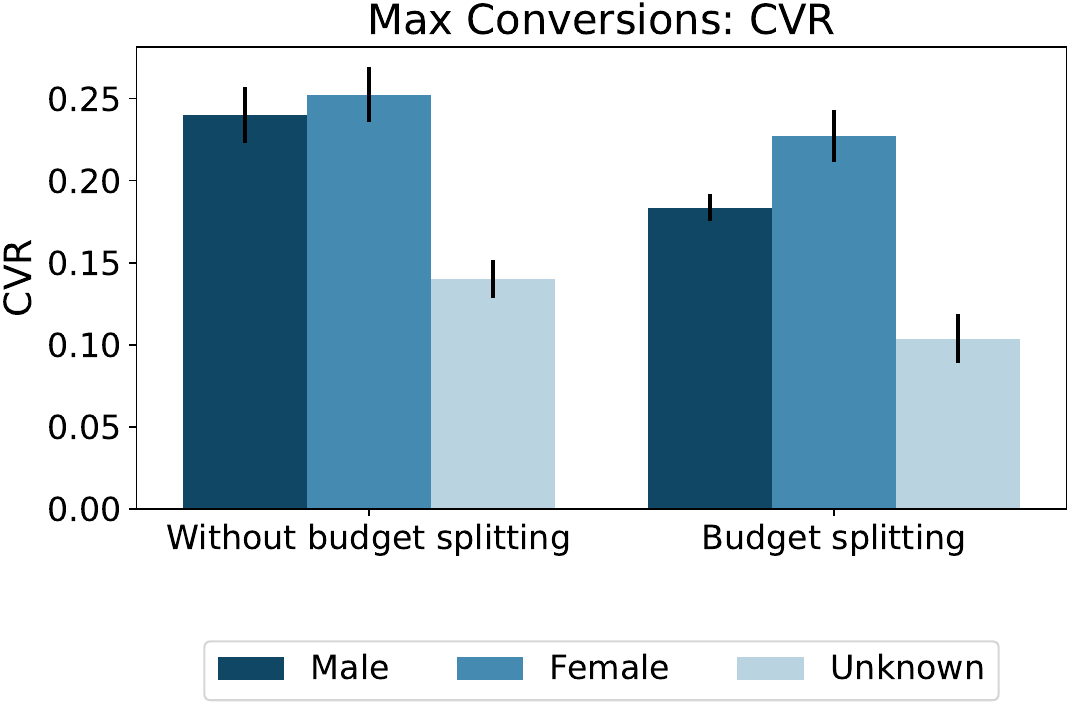}
    \caption{}
  \end{subfigure}
  \caption{
  Average weekly (a) Click-through rate (CTR) for Max Clicks campaign before and after budget split, (b) Impressions to Conversion rate (CVR) for Max Conversions campaign with and without (before) the budget split.
  Standard error bars are shown.
  }
  \Description{Bar plot with standard error bars showing the average weekly CTR for each targeting approach.}
  \label{fig: cvr and ctr}
\end{figure}

\begin{figure}[ht]
  \centering
  \includegraphics[width=.8\textwidth]{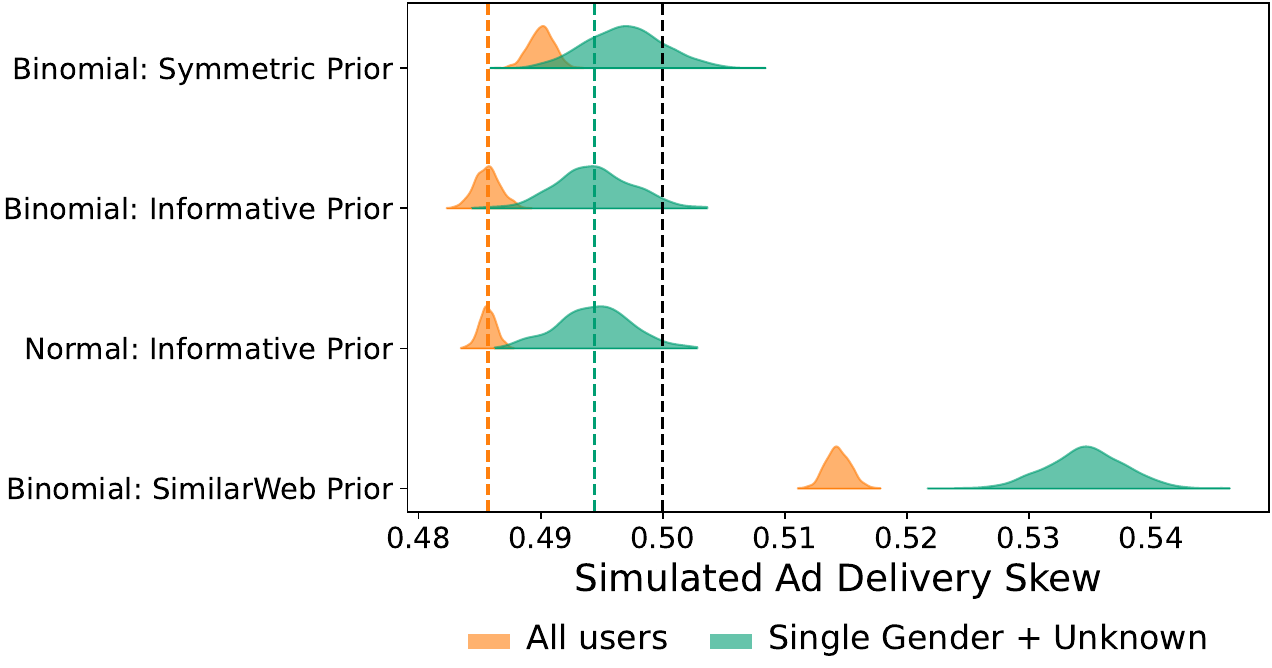}
  \caption{Distributions of ad delivery skew for Max Conversions campaign under four sampling distributions (N=1,000) of male individuals in the unknown user pool. Vertical lines depict the estimated skew (calculated without unknown users) for All Users (orange) and Single Gender + Unknown Users (green) when calculated with only Google-labeled male and female users.}
  \Description{Distributions of ad delivery skew for the max conversions campaign under simulated populations of unknown users.}
  \label{fig: max conversions simulations}
\end{figure}

\begin{figure}[ht]
  \centering
  \includegraphics[width=.8\textwidth]{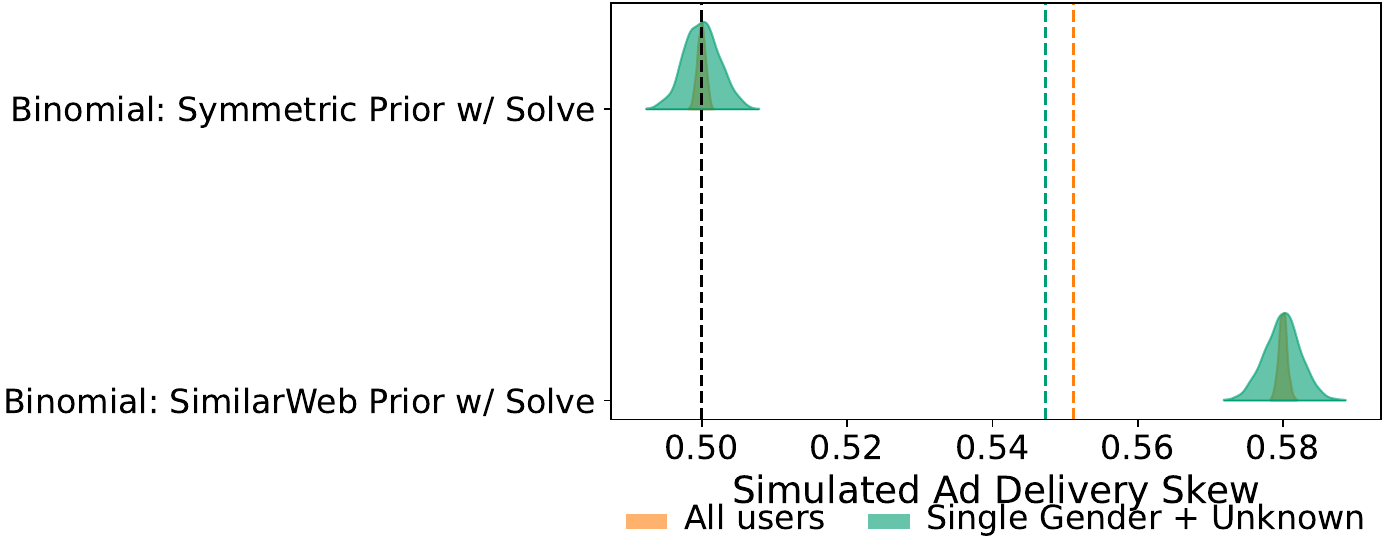}
  \caption{Distributions of ad delivery skew for Max Clicks campaign under two sampling distributions (N=1,000) of male individuals in the unknown user pool. Vertical lines depict the estimated skew (calculated without unknown users) for All Users (orange) and Single Gender + Unknown Users (green) when calculated with only Google-labeled male and female users.
  We observe that simulations of ad delivery skew that solve for the complete distribution reflect simulation parameters.}
  \Description{Distributions of max clicks campaign skew under simulations of unknown user populations, such that the simulated populations are generated by considering the complete distribution (including known users).}
  \label{fig: max clicks simulations with solve}
\end{figure}

\end{document}